# Tomographic diffractometry of laser-induced plasma formations

*Ivan Ostrovsky[1,2], Gilad Hurvitz[3], Eli Bograd[1,2], Eli Flaxer[1,4], Soumitra Hazra[1,2],*

*and Sharly Fleischer[1,2]*

[1] School of Chemistry, Raymond and Beverly Sackler Faculty of Exact Sciences, Tel Aviv University, Tel Aviv 6997801, Israel.
[2] Tel Aviv University Center for Light-Matter Interaction, Tel Aviv University, Tel Aviv 6997801, Israel.
[3] Applied Physics Division, Soreq NRC, Yavne 81800, Israel
[4] AFEKA - Tel-Aviv Academic College of Engineering, 69107 Tel-Aviv, Israel.

## Abstract

A sensitive optical diffractometry method is developed and utilized for advanced tomography of laser-induced air plasma formations. Using transverse diffractometry and Supergaussian plasma distribution modelling we extract the main parameters of the plasma being the plasma density, width and shape with $20 \mu m$ spatial resolution throughout the plasma formation. The experimentally recorded diffraction patterns fitted by the Supergaussian plasma model are found to capture unprecedentedly delicate traits in the evolution of the plasma from its effective birth and on. Key features in the spatial evolution of the plasma such as the 'escape position', the 'turning point' and the refocusing dynamics of the beam are identified and explored in details. Our work provides experimental and theoretical access into the highly nonlinear dynamics of laser-induced air plasma.

## Introduction

Plasma, first described by W. Crooks as "radiant matter" in 1879 [1] is the state of matter composed of a gas of ions and free electrons and is a central pillar of physics since the early experiments of Langmuir and co. in the late 1920s [2]. The majority of laboratory plasma research relies on electrically and light-induced plasmas, created by intense fields covering a broad range of frequencies such as discharge under constant or pulsed DC fields, RF and microwave [3], electron beams and high intensity lasers [4] to name a few. Such sources provide partially ionized plasmas that undergo recombination processes often involving multiple chemical reactions as well as spatial diffusion and temperature modifications, together giving rise to intricate spatial and temporal plasma dynamics. World-celebrated advents in chirped-pulse amplification (CPA) enabled table-top generation of ultrashort laser pulses with intensities exceeding $10^{14} \, W/cm^2$ at the focus [5] and create partially ionized air plasma on femtosecond (fs) time scales. Laser filamentation, i.e. the creation of elongated plasma channels in air by intense fs pulses has attracted vast scientific interest over the last decades [4,6,7] fueled by various prospected applications [8–11] including taming lightening breakdown with lasers [12], remote sensing of chemicals in the atmosphere [13], light waveguiding through dense clouds [14], defense applications [15] and even weather control [16,17].

The propagation of intense fs pulses in a medium (typically air) exhibits various nonlinear optical phenomena and has sparked a broad interest and vast theoretical and experimental efforts. Very briefly, Kerr-nonlinearity gives rise to self-focusing of the pulse [7,18] that is counteracted by plasma defocusing [19,20], interweaved with dispersive effects in the bare and partially ionized air, plasma absorption and avalanche ionization processes that together give rise to intricate spatial and temporal evolution of the pulse and result in complex spatio-temporal plasma formation [19,21–24].

**Here we** implement a transverse ultrafast optical diffractometry technique for spatial characterization of laser-induced plasma entities. From the measured diffraction patterns we analytically extract a spatial phase induced by the plasma, and by fitting the latter to a Super-Gaussian (SG) plasma distribution, obtain the local density, shape and size of thin plasma slabs (~$20\mu m$ resolution) along the pulse propagation direction. The high experimental sensitivity provides access to unprecedentedly weak plasma entities induced by as low as few tens of $\mu J$ pulse energies, revealing the transition between linear and nonlinear regimes of pulse propagation. The experimental results validate and are complemented by theoretical simulations that are analyzed in details. The SG plasma model utilized here is found to be crucial for understanding several universal features in the evolution of laser-induced plasmas.

***The paper is organized as follows:*** Section 1 describes the motivations of this work and the existing methodologies for ultrafast plasma characterization. Section 2 describes the experimental technique and the analysis strategy. Section 3 presents the experimental and theoretical results that are analyzed and discussed in section 4.

### 1) State-of-the art and motivation

Intense laser pulses exceeding the critical power of $2.4 GW$ undergo self-focusing as they propagate in air [7,25] which leads to plasma formation and the collapse of the pulse after few tens of meters [18]. More compact laboratory experiments can be realized using external focusing elements such as lenses or curved mirrors [26–28]. With focal length $f < 750 mm$ the geometrical focusing imposed by the lens overshadows the self-focusing effect [29] and the high plasma densities impart strong defocusing on the beam, thwarting multiple focusing-defocusing cycles and rendering the plasma channel termination in vicinity of the geometrical focal point [26]. Nevertheless, the spatial and temporal characteristics of lens-assisted plasmas exhibit intriguing dynamics, the study of which is crucial for unveiling the underlying physical and chemical evolution of the plasma, and for unleashing its potential for a variety of applications.

Various methodologies have been implemented for studying laser-induced plasma over the years. Electrical conductivity measurements [30], microwave diagnostics [31], and Schlieren photography [32] provide information on the density of long-lived plasma entities, however lack the temporal resolution required for ultrashort-laser induced plasmas. Experimental signatures such as the photoemission upon plasma recombination [33,34], spatial and spectral characterization of the pulse following self-phase modulation [35–37] imaging of the pulse wavefront [37,38] and detection of THz radiation emission that is correlated with the frequency of plasma oscillations [39] provide complementary information about the intricate plasma dynamics and the nonlinear processes experienced by the short laser pulse. Direct probing techniques rely primarily on the refractive index of the plasma as the key observable. Wave-front optical interferometry [40,41] monitors the distortions invoked on the wavefront of a short laser pulse upon traversing the plasma filament at a small angle. Plasma-enhanced third harmonic generation relies on the 3[rd] order susceptibility of the plasma $\chi^{(3)}$ for probing the spatial distribution and temporal evolution of the plasma [42,43]. Time-resolved optical diffractometry uses the diffraction of a weak probe pulse off the plasma entity and can be realized both in a longitudinal [26,44,45] and transverse [30,46–49] geometry, the latter is used in this work.

### 1.1) The problem

Optical diffractometry and interferometry consider the plasma as a spatially localized phase object imposed on a probe pulse traversing it. This phase emanates from the change in refractive index of the plasma with respect to the native surrounding medium (typically atmospheric air) and on the spatial distribution of the plasma entity. Whether it is the interference fringe pattern or the far-field diffraction pattern, the effective phase imposed on the probe is integrated across the plasma cross section (with a few tens of $\mu m$ typical diameter), thus casting a 1D phase object over a traversing probe field. To retrieve the plasma density and spatial distribution from the experimentally obtained 1D phase object, a model for the plasma cross-section at the intersection with the probe must be assumed. Indeed, the plasma cross section has been conveniently modeled as a Gaussian electron density [30,40,46,47,49–52] or as a uniform (cylindrical) density [26,44,45,48,53] where the 1D phase imposed by the plasma is obtained via integration along the probe propagation direction. In this work we study plasma formations induced with varying pulse intensities that span the range between the purely Gaussian model (valid for very low intensity laser pulses) and uniform density (valid for high intensity laser pulses) distributions. Here we model the plasma density as a SG distribution and show that it captures the fine details of the diffraction patterns and provides complete information on the plasma density, size and shape. More importantly, the SG model inherently bridges between the typically used Gaussian and uniform distribution models in a smooth and continuous way (see comparison of the pure Gaussian, SG and uniform density models in supplementary information section 1). The SG order extracted from the experimental results is found to be a key parameter for plasma characterization and its spatial evolution. An elegant analytic methodology is developed and utilized for extracting the plasma-induced 1D phase as will be described hereafter.

## 2) Experimental system

The experimental setup used in this work is depicted in Figs.1a,b. A short laser pulse (120fs duration, $\lambda_c = 800nm$) from a chirped pulse amplifier (CPA) is split in two parts to yield a strong pump pulse (~1mJ) and a weak probe pulse (few tens of $\mu J$). The pump pulse is routed through an attenuator ($\lambda/2$ waveplate + thin film polarizer (P1)) and is focused by a lens (L1, f=500mm) mounted on a computer-controlled stage (S1) to vary the position of the plasma. The weak probe pulse is routed through a computer-controlled 150mm delay stage enabling up to 1ns delay between the probe and the pump at their point of intersection. The probe is focused by a cylindrical lens (CL, f=100mm) to form a thin (~20$\mu m$ at the focus) and long (~12$mm$) light-line that is diffracted off the plasma at 90° onto a photo-detector (PD) equipped with a narrow (20$\mu m$) slit, the position of which is scanned in the y-axis direction (Fig.1b) to scan across and detect the diffraction pattern. The long (12$mm$) probe pulse ensures a uniform intensity at the intersection region with the ($< 100\ \mu m$) plasma formation. The output of the PD is sampled by a lock-in amplifier at the pump frequency that is modulated at subharmonic (500Hz) repetition frequency of the probe (1kHz) using a mechanical chopper wheel. This configuration provides selective detection of the plasma induced diffraction pattern with high sensitivity and noise rejection and enables experimental monitoring of unprecedentedly weak plasma entities induced by as low as ~50 $\mu J$ pulse energies. For high intensity plasma generation further reduction of the background light emitted upon plasma recombination and white-light scattering is provided by bandpass filter ($800 \pm 20nm$, FBH800-40 by Thorlabs) and differential modulation detection with two chopper wheels at $500Hz$ and $333Hz$ and detection at their difference frequency of $166Hz$ [54,55]. Fig.1c shows an example of a diffraction pattern obtained in our experiment (blue curve) and its numerical SG fit (orange curve). For additional diffraction examples and their SG fit, see supplementary information section 2.

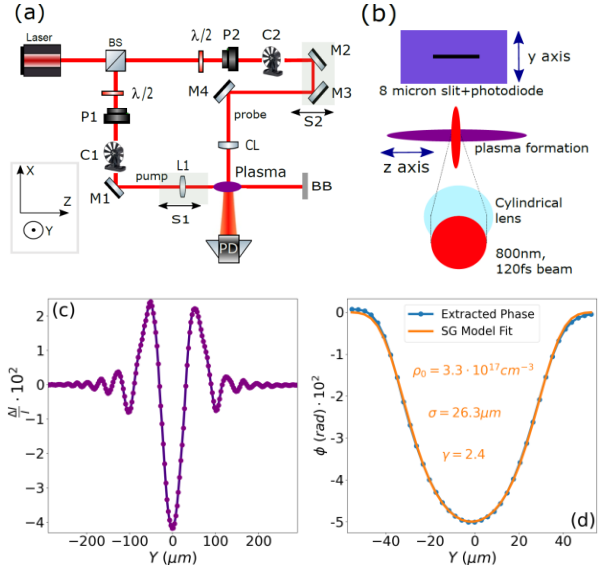

Figure 1 (a) A schematic drawing of the experimental setup. (b) Enlarged view of the interaction region between the probe and plasma. The probe beam (red) is focused onto the plasma channel by a cylindrical lens and diffracts onto a photodiode (PD) equipped with a narrow $20\mu m$ slit. By translating lens L1 along the pump propagation direction we vary the position at which the probe intersects with the plasma channel. The diffraction pattern is recorded by scanning the position of the PD in the y-direction. (c) Example of an experimental diffraction pattern from plasma ($640\mu J\ pulse\ energy$, $500mm$ lens, $z = 496mm$, i.e. $4mm$ prior to the geometric focal point of the lens). The probe was delayed $25ps$ past plasma generation and the PD positioned $5.3mm$ after the plasma. (d) Experimental phase (blue points) and SG fit (orange curve) from which the SG parameters are extracted.

## Diffraction pattern analysis: plasma phase extraction and SG model fitting

For each position along the plasma channel (set by the position of S1) and pump-probe delay of $25ps$ (set by S2) we obtain an intensity modulation pattern like that depicted by the purple curve in Fig.1c. The intensity modulation pattern $\frac{\Delta I}{I}$ corresponds to the difference between the probe intensity on the screen (PD), with and without plasma (marked $I_{plasma}$ and $I_0$ respectively): $\frac{\Delta I}{I} = \frac{I_{plasma} - I_0}{I_0}$.

The plasma cross-section is taken as a spatial charge distribution $\rho_e(x,y)$, the refraction index of which is given by $n(x,y) \cong n_0 - \frac{\rho_e(x,y)}{2\rho_c}$ where $\rho_c = 1.7 \cdot 10^{21} cm^{-3}$ is the critical plasma density at the probe wavelength ($\lambda = 800nm$) [4] and $n_0$ is the refractive index of the ambient laboratory atmosphere. To calculate the phase difference imposed on the probe upon propagation across the plasma, we integrate over the $x$–dimension of $n(x,y)$ and multiply by the probe wavenumber ($k_{800nm}$):

$$(1)\ \Delta\varphi(y) = k_{800} \int_{-\infty}^{\infty} [n(x,y) - n_0]\, dx = -\frac{2\pi}{\lambda_{probe}} \frac{1}{2\rho_{crit}} \int_{-\infty}^{\infty} \rho_e(x,y)\, dx\ .$$

Different from the previously used Gaussian and uniform cylinder models, here we use the SG distribution: $\rho_e(x,y) = \rho_0 e^{-\left(\frac{x^2+y^2}{2\sigma^2}\right)^{\gamma}}$ where $\rho_0$ is the charge density at the center of the plasma cross-section, $\sigma$ and $\gamma$ are the variance and SG order respectively. Note that the SG order $\gamma$ spans the entire range of plasma distributions bound between the two above-mentioned models, where $\gamma = 1$ is a pure

Gaussian and $\gamma \gg 1$ approaches the uniform distribution model (see supplementary information section 1 Fig.S2). The 1D integration over the SG distribution however, requires numerical evaluation.

Our experimental diffractometry parameters are well within the Fresnel diffraction approximation as indicated by the Fresnel number $N_F = \frac{a^2}{D\lambda} \in [0.02, 2.5]$ with $a$- object size [10 $\mu m$ -100 $\mu m$], $D = 5.3mm$ is the distance from the plasma object to the screen, and $\lambda = 800nm$ for the probe wavelength of the probe. These parameters satisfy the Fresnel accuracy condition laid in [56] equation 4-18.

In order to extract the experimental plasma phase $\Delta\varphi(y)$ from the measured diffraction pattern we utilize a linear algorithm reminiscent that of ref. [57] described hereafter:

The Fresnel diffraction pattern is given by the convolution of the probe field at the object position and the Fresnel phase: $h_z(y) = \frac{e^{ikz}}{i\lambda z} e^{i\frac{k}{2z}y^2}$.

$E_p^z(y)$ stands for the diffraction field at the screen position. $E_0^0(y)$ and $E_0^p(y)$ are the probe field at the plasma plane ($z = 0$) with $E_0^0$ as the bare probe field and $E_0^p$- the probe imposed with the plasma phase: $E_0^p(y) = E_0^0(y) \cdot e^{i\Delta\varphi(y)}$. For sufficiently small plasma phase $\Delta\varphi(y)$ such as in dilute laser-induced air plasmas ($\rho_0 < 10^{19} cm^{-3}$) and typical plasma diameter~$100\mu m$ we use the small angle approximation to write: $e^{i\Delta\varphi(y)} \approx 1 + i\Delta\varphi(y)$.

Let us write the diffracted field at the plane of the screen ($z$):
$E_p^z = E_0^p * h_z$ where ' * ' stands for convolution.
By the convolution theorem, we can calculate the expected intensity at the PD:

(2) $I_z^0 = |E_0^0 * h_z|^2 = |\mathcal{F}^{-1}\{\mathcal{F}\{E_0^0\} \cdot \mathcal{F}\{h_z\}\}|^2$

(3) $I_z^p = |\mathcal{F}^{-1}\{\mathcal{F}\{E_0^0 \cdot e^{i\Delta\varphi}\} \cdot \mathcal{F}\{h_z\}\}|^2$

Where $\mathcal{F}$ and $\mathcal{F}^{-1}$ stand for Fourier and inverse Fourier transform respectively.

Substituting the phase as noted above and assigning $H_z(f_y) \equiv \mathcal{F}\{h_z(y)\} = e^{ikz} e^{-i\pi\lambda z f_y^2}$ we get:

(4) $I_z^p(y,z) \approx |\mathcal{F}^{-1}\{\mathcal{F}\{E_0^0\} \cdot H_z\} + \mathcal{F}^{-1}\{\mathcal{F}\{E_0^0 \cdot i\Delta\varphi\} \cdot H_z\}|^2 =$
$= |E_z^0|^2 + |\mathcal{F}^{-1}\{\mathcal{F}\{E_0^0 \cdot i\Delta\varphi\} \cdot H_z\}|^2 + E_z^{0*} \cdot \mathcal{F}^{-1}\{\mathcal{F}\{E_0^0 \cdot i\Delta\varphi\} \cdot H_z\} + E_z^0 \cdot \mathcal{F}^{-1}\{\mathcal{F}\{E_0^0 \cdot i\Delta\varphi\} \cdot H_z\}^*$

We discard of the 2nd term in the right hand side of eq.(4) as it is negligibly small with respect to the other terms owing to the inherently small $\Delta\varphi$ to the power of 2. Furthermore, within the narrow range scanned along the $y$ − axis ($< 400 \mu m$), the intensity of the probe beam is practically constant in the central ~$400 \mu m$ of a ~$12 mm$ long light sheet. Thus, we can safely approximate a constant probe amplitude $E_z^0 \approx \sqrt{I_z^0}$. We note that owing to the short propagation distance from plasma position to the detection plane together with the wide aperture of the slit in the z-axis direction, the probe light is captured in full, hence $E_0^0 \approx E_z^0$. As a result:

$I_z^p = I_z^0 + E_z^0 \cdot (\mathcal{F}^{-1}\{\mathcal{F}\{E_0^0 \cdot i\Delta\varphi\} \cdot H_z\} + \mathcal{F}^{-1}\{\mathcal{F}\{E_0^0 \cdot i\Delta\varphi\} \cdot H_z\}^*)$

$\frac{\Delta I}{I} \equiv \frac{I_z^p - I_z^0}{I_z^0} = \mathcal{F}^{-1}\{\mathcal{F}\{i\Delta\varphi\} \cdot H_z\} + \mathcal{F}^{-1}\{\mathcal{F}\{i\Delta\varphi\} \cdot H_z\}^*$

Rearrangement of the above equation gives an analytical expression for the phase

(5) $\Delta\varphi = \mathcal{F}^{-1}\left\{\frac{\mathcal{F}\{\Delta I/I\}}{2\sin(\pi\lambda z f_y^2 - kz)}\right\}$

Equipped with a closed analytic procedure to retrieve $\Delta\varphi$, we perform a numerical fit of our experimentally recorded diffraction patterns (as shown in Fig. 1d) to eq. (1) and extract the parameters of the SG plasma distributon. To find the correct set of plasma parameters $(\rho_0, \sigma, \gamma)$ that give rise to the experimentally obtained phase object we numerically perform SG integrals (as in eq.1) that seed a non-linear least squares curve fitting algorithm.

### 3) Experimental and Theoretical results

Figure 2 depicts the SG parameters extracted from a set of plasma formations differing in pump pulse energy ranging from $100\mu J$ to $860\mu J$ (color-coded). The pump beam is focused by an $f = 500mm$ lens, the geometrical focal point is marked $'500'$ at the z-axis (for methods used to determine the exact focal position of the lens see supplementary information section 3). The probe is timed to intersect the plasma at $\Delta t = 25ps$ past its creation to avoid temporal overlap between the pump and probe pulses, and to avoid significant recombination of the plasma.

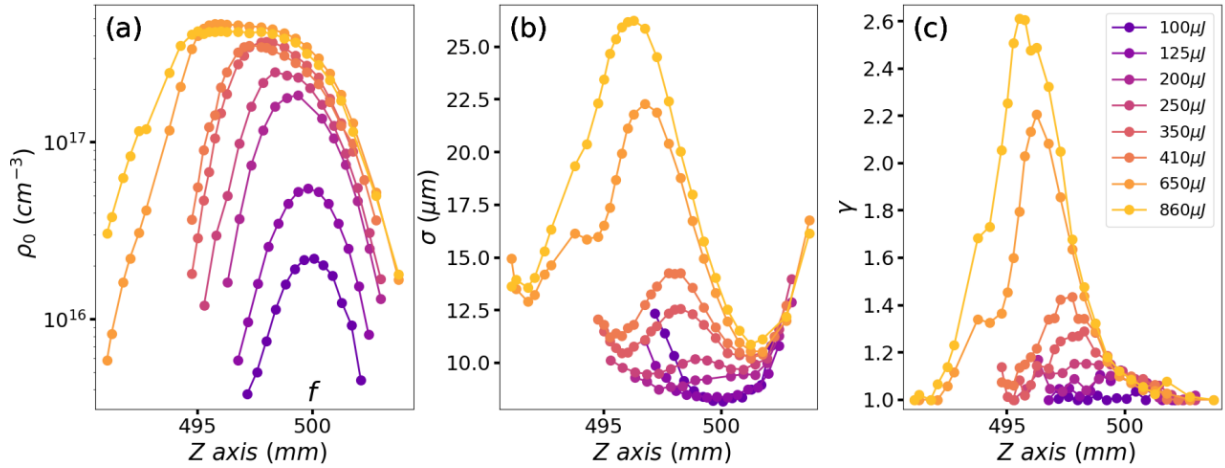

Figure 2: Plasma density, σ and γ along the pulse propagation axis ($Z$) for different pump pulse energies (color-coded).

For each pump energy we scan a range along the $z$-direction in which the diffraction patterns are detectable and extract the spatially resolved SG plasma parameters: $\rho_0$, $\sigma$ and $\gamma$ shown in panels a,b,c respectively. At the lowest pump energy ($100\mu J$, purple curve) the plasma density peaks at the geometrical focus of the lens. At this pulse energy, the propagation of the pulse is practically unaffected by the created plasma ($\rho_0 \sim 10^{16}\ cm^{-3}$) and the Gaussian shape is retained as evident from the $\sigma$ and $\gamma = 1$ values in the $497mm - 502mm$ range. As the pump energy increases, plasma defocusing starts playing an increasingly important role, rendering the beam propagation nonlinear. This manifests as a gradual shift in plasma density toward the direction of the lens (smaller $z$ values) and by the increasing $\gamma$ values of the plasma shape. Note that higher pump energies induce denser plasma detectable at z-positions preceding the focal point and give rise to increasing asymmetry of the plasma distribution density around the geometrical focal position. This is also observed by the asymmetry in σ and γ parameters with the latter gradually increasing up to $\gamma \sim 2.6$ for $860\mu J$ pump energy. Intriguingly, all of the plasma formations shown in Fig.3 are found to terminate around the geometrical focal point and coincide at $z \sim 502.5mm$ where the pure Gaussian beam shape is retrieved with $\sigma \sim 12.5\mu m\ and\ \gamma = 1$. Saturation of the plasma density around $\rho_0 \sim 4 \cdot 10^{17} cm^{-3}$ is observed above $650\ \mu J$ owing to intensity clamping [58], in excellent agreement with the saturation values obtained by the simulations shown in Fig.3a.

The simulations of Fig.3 were conducted in a similar approach to ref. [20,29]. Briefly, we propagate the non-linear Schrödinger equation (NLSE) with nonlinear terms standing for the plasma defocusing, depletion, time evolution and self-focusing and record the generated plasma as well as the pulse parameters throughout the entire evolution. For a detailed description of the simulation method, see Supplementary Information section 4 which includes ref [59]).

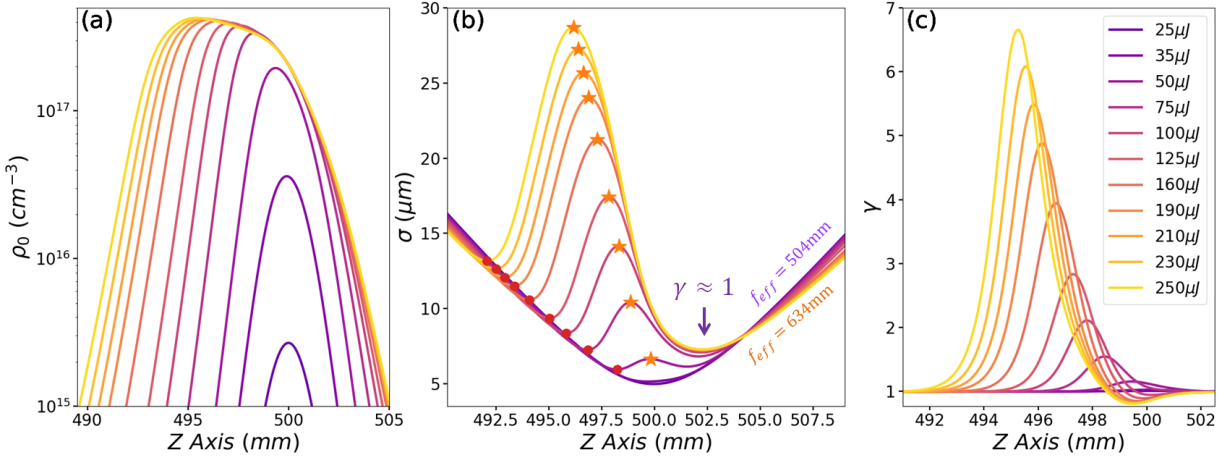

Figure 3:(a,b,c) Simulated Plasma densities $\rho_0$, σ and γ as a function of position along the beam propagation axis ($z$), similarly to Figure (2). The experimental and simulated data are in excellent agreement. To obtain the best agreement with experimental results, the pulse energies used in simulations are ~3.4 smaller than the experimental pulse energies (see text). The red points in Fig.(b) mark the 'escape positions' as calculated from eq.(7) in section 4.1.

Before we proceed with the discussion we note the discrepancy in pulse energies used in fitting the simulated data to the experimental data with a factor of ~3.4 between the two. The same discrepancy was reported previously in [26] [29], the reason for which may be attributed to the convention used for energy-to-intensity transformation and requires further exploration. With this factor included, the transition from linear to non-linear propagation is observed between $35\,\mu$J~50µJ in the simulation and agrees with the experimental range of 125µJ~200µJ. We wish to stress the excellent agreement between the simulated and experimental results for all three parameters $\rho_0, \sigma, \gamma$. The difference between the $\gamma$ values in Fig.3c and Fig.2c results from our decision to plot the plasma parameters as extracted from the raw experimental data (without post processing). This difference in $\gamma$ attributed to the convolution of the diffraction pattern with the slit that is mounted on the PD (essentially a window function of 20µ$m$ width). We have verified that while the $\rho_0$ and $\sigma$ parameters are hardly affected by the above, the extraction of large $\gamma$ values is more sensitive to the exact curvature of the peaks of the diffraction pattern. Accounting for the latter we have found that $\gamma = 6$ and $\gamma = 2.1$ reduce to $\gamma$ ~2.6 and $\gamma$ ~1.34 respectively following convolution of the raw diffraction patterns with a 20µ$m$ window, in excellent correspondence with the experimental values. For the 'raw' and convoluted diffraction patterns and their fitted SG parameters see supplementary information section 5.

## Analysis of key features in the longitudinal plasma evolution

From the results of Fig.2 and Fig.3 we identify several key features that are shared by all of the plasma formations in our range of pump energies. These are presented and analyzed in the remaining of this paper.

**4.1) Escape position –** In each of the curves depicted in Figs. 2,3, one finds a specific position along the propagation axis where the beam strongly diverges from the linear, geometric evolution imposed by the lens, as evident by the rapid increase in the width ($\sigma$) of its legacy plasma. At these 'escape positions' the wavefront curvature of the beam switches sign from negative (focusing) to positive (defocusing), crossing through zero. In order to evaluate the escape positions ($z_{esc}$) we calculate the parabolic phase coefficient of the beam's wavefront throughout the $z$-axis by taking the second derivative with respect to $r$ and identify the zero-crossing position (where the parabolic phase changes sign). The overall wavefront phase is contributed by the following terms:

**Geometric phase** of the lens: $\phi_{geometric}(r, z_{esc}) = \dfrac{r^2}{2z_{esc}\left[1+\left(\frac{z_R}{z_{esc}}\right)^2\right]}$

where $z_R$ is the Rayleigh range of the beam

**Plasma phase** (Defocusing): $\phi_{plasma}(r,z) = \int_{-f}^{z_{esc}} -\dfrac{\rho_{nt}\tau\sigma_k}{2\rho_c} I^8(r,z)\, dz$

where $I(r,z) = I_0 \left(\dfrac{\omega_0}{\omega(z)}\right)^2 e^{-\frac{2r^2}{\omega^2(z)}}$, $\rho_{nt} = 5.4 \cdot 10^{18} cm^{-3}$ is the density of neutral oxygen in air, $\tau = 120 fs$ is the pulse length, $\rho_c = 1.7 \cdot 10^{21} cm^{-3}$ is the critical plasma density and $\sigma_k = 2.81 \cdot 10^{-96} W^{-8} cm^{16} s^{-1}$ is the cross-section for multi-photon ionization [29].

**Kerr phase** (Focusing): $\phi_{Kerr} = \int_{-f}^{z_{esc}} n_2 I(r,z)\, dz$ where $n_2 = 3.2 \cdot 10^{-19} cm^2 \cdot W^{-1}$ is the Kerr coefficient for air [4].

Different from $\phi_{geometric}(r,z)$ which is predetermined by the lens and known at each $z$-position, the nonlinear $\phi_{plasma}$ and $\phi_{Kerr}$ are accumulated throughout the propagation hence must be integrated over the entire propagation length. The condition for $z_{esc}$ is thus:

$$(6) \quad \dfrac{d^2}{dr^2}\left\{\phi_{geometric}(r,z_{esc}) + \int_{-f}^{z_{esc}} \phi_{plasma}(r,z)dz + \int_{-f}^{z_{esc}} \phi_{Kerr}(r,z)dz\right\} = 0$$

Equation 6 can be solved numerically around $r \sim 0$ by Taylor expansion of the exponential terms, i.e. at the vicinity of the center of the beam which is primarily susceptible to the nonlinear phase modifications owing to the high peak intensity at the center. Following rearrangement the equation takes the form:

$$(7) \quad \dfrac{1}{z_{esc}\left[1+\left(\frac{z_R}{z_{esc}}\right)^2\right]} + \dfrac{16\rho_{nt}\tau\sigma_k}{\rho_c}\dfrac{I_0^8}{w_0^2}\int_{-f}^{z_{esc}}\left[1+\left(\frac{z}{z_R}\right)^2\right]^{-9} dz - \dfrac{4n_2 I_0}{w_0^2}\int_{-f}^{z_{esc}}\left[1+\left(\frac{z}{z_R}\right)^2\right]^{-2} dz = 0$$

The numerically extracted $z_{esc}$ for the different pulse energies are marked by the red points in Fig.3b, and are in very good agreement with the full NLSE calculations as evident by their close proximity to the first minima of $\sigma(z)$. Slight deviations of the calculated $z_{esc}$ using eq. (7) from the full NLSE are expected since the former inherently discards any propagation effects that are experienced by the beam and accounted for in the latter. The pulse intensity at the 'escape positions' was found to be $\sim 2.7 \cdot 10^{13} \dfrac{W}{cm^2}$ across all of the pulse energies simulated in Fig.4.

**4.2) Turning point and refocusing:** As the beam diverges from its lens-imposed focusing geometry, it defocuses quite rapidly as evident by the increasing $\sigma$ and $\gamma$ parameters extracted from its legacy plasma formation. The gradual increase in beam area ($\propto \sigma^2$) results in decreased pulse intensity thus expected to eventually suppress further plasma generation. Nevertheless, as evident from the experimental and numerical results, substantial defocusing of the beam is observed e.g. by the yellow curve (250μJ in Fig.4b) as $\sigma$ increases from $\sim 12.5\mu m$ to $\sim 26\mu m$ over a propagation distance of $\sim 4mm$. To delve into the underlying beam dynamics one must consider the gradual reshaping of the beam as manifested by the increasing $\gamma$ (the SG order parameter) as follows: The highest intensity of the initially Gaussian beam (in vicinity of the escape position) is the first to generate plasma and experience defocusing. This results in the gradual 'flattening' of the beam intensity distribution (starting from the center and progressing radially outwards) which is directly associated with the increasing $\gamma$ parameter. Note that different from the 'auto-enhancing' nature of the Kerr effect, plasma defocusing and intensity flattening (clamping) compromises further plasma defocusing as it diminishes the gradient of the plasma density across the beam. The latter, together with the gradually reducing intensity noted above limit the extent of plasma formation and beam defocusing. A maximal value in both $\sigma$ (marked by orange stars in Fig.3b) and $\gamma$ is thus observed shortly after and referred to here as the 'turning point'.

Figure 4 depicts a series of snapshots showing the intensity distribution and wavefront phase across the beam diameter between the escape position and the turning point. These snapshots were extracted from a simplified version of the NLSE simulation where temporal effects are discarded (typically referred to as 'frozen time' [60] [61], see supplementary information section 4) however verified to be in good agreement with the full NLSE dynamics of Fig.3. The evolution of the intensity distribution (red curves in Fig.4) from pure Gaussian at the escape position to a maximal SG order around the turning point are clearly observed. The evolution of the wavefront phase is depicted by the pale blue curves, showing gradual flattening of its initial parabolic phase toward complete erasure of the phase gradient (flat phase across the beam) reached at the turning point.

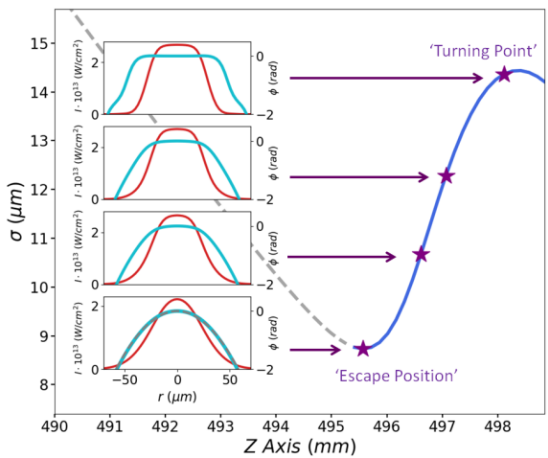

Figure 4: Radial beam intensity and phase evolution vs propagation distance for simulated $100\mu J$ pulse energy. The phase is erased in the viscinity of the $\sigma$ peak. 'Frozen time' approximation is used for simplification. The intensity and the phase of the beam are marked in red and cyan respectively. The geometrical phase is shown in purple for comparison.

Surprisingly, despite its utterly flattened phase, the beam reverts back to a Gaussian intensity distribution ($\gamma \rightarrow 1$) shortly after the turning point. This behavior is dictated by the natural evolution of SG beams that tend to focus and self-reshape to Gaussian intensity profile under free propagation [62]. In our case, the plasma generated throughout this process serves to restrict the beam intensity and results in monotonic decrease in intensity and a moderate transition toward the Gaussian profile. As observed in the experimental and theoretical results of Figs.2,3, all the plasma formations terminate around the same position (marked in Fig.3b by the purple arrow positioned at $\gamma\sim1$) after which pure Gaussian beam propagation (defocusing) is observed. Note that while the simulated beams converge back to Gaussian profile, $\gamma$ slightly overshoots the value of 1 from below (minimal value of $\gamma\sim0.8$ reached for the simulated $250\mu J$ pulse in Fig.3c). The latter however is not captured by the experimental results of Fig.2c. While the convolution with the window function of the slit shifts $\gamma = 0.8$ to $\gamma\sim0.9$ (see Supplementary Information Section 5) this discrepancy may possibly hint on additional effects that may be discarded from the current NLSE simulation, motivating further exploration and improvement of both the simulated and experimental techniques.

**4.3) <u>Plasma as an effective lens</u>**: A detailed inspection of the beam diameter after the plasma formation reveals the effect of the accumulated plasma defocusing on the divergence of the beam. Pulses with higher initial energies show smaller divergence after regaining their Gaussian shape as readily observed on the right-hand side of Fig.3b. The opposing contributions of the geometrical lens and the accumulated plasma can be considered as an 'effective lens'. This is done by taking the beam waist $w(z_{1,2})$ at two positions $z_{1,2} \gg z_R$ (where $z_R = \frac{\pi w_0^2}{\lambda}$ is the Rayleigh length) and using the known relations for Gaussian beams: $w(z) = w_0 \sqrt{1 + \left(\frac{z}{z_R}\right)^2} \approx \frac{z \cdot \lambda}{\pi \cdot w_0}$

Taking the derivative with respect to $z$: $w' = \frac{dw(z)}{dz} = \frac{\lambda}{\pi \cdot w_0}$ we obtain $f_{eff} = \frac{2 \cdot D}{\lambda \cdot w'}$.

With $w' = \frac{w(z_2) - w(z_1)}{z_2 - z_1}$ extracted from the simulation, $D = 4.5mm$ - the beam diameter at the lens position and $\lambda = 800nm$ we calculate $f_{eff} = 504mm$ and $634mm$ for the simulated $25\mu J$ and $250\mu J$ pulse energies respectively. Note that $f_{eff} > f_{lens}^{500mm}$ as expected.

## **Summary**

In this work we have demonstrated advanced tomographic characterization of laser-induced air plasma formations. The uniquely sensitive experimental setup together with the SG plasma modeling and supporting theoretical simulations provide dramatically improved accessibility into the underlying phenomena that dictate the longitudinal plasma evolution. We have followed this evolution starting from the effective birth of the plasma (where the beam propagation is unaffected by its generated plasma) and far beyond into the nonlinear propagation regime - where the interplay between the beam and plasma governs the dynamics. The presented SG plasma model unifies and bridges between the previously used models (pure Gaussian and uniform density) in a smooth and continuous manner, and as manifested by its excellent fit to the experimental diffractometry results. Moreover, the evolution of the SG order parameter is crucial for understanding the key features that govern the spatial evolution of the beam and plasma described in section 4. The experimental methodology and advanced analysis presented here enables unprecedented quantification of laser induced plasma formations. Thus our work pave new path for addressing various challenges in the determination of plasma properties and parameters as well as for various applications of laser-induced plasma. We note that while this work primarily focused on the spatial evolution of the plasma, our setup enables monitoring of the temporal evolution of the plasma as well. This will be described in a forthcoming publication.

**Acknowledgment:** The authors acknowledge the support of Pazy Foundation grant no. 5100057425 and the Israeli Science Foundation grant no. 1856/22. We thank Prof. Guy Cohen (TAU) for providing computational resources for our simulations.

# Supplementary Information –

# Tomographic diffractometry of laser-induced plasma formations

*Ivan Ostrovsky[1,2], Gilad Hurvitz[3], Eli Bograd[1,2], Eli Flaxer[1,4], Soumitra Hazra[1,2],*

*and Sharly Fleischer[1,2]*

[1] School of Chemistry, Raymond and Beverly Sackler Faculty of Exact Sciences
[2] Tel Aviv University Center for Light-Matter Interaction, Tel Aviv University, Tel Aviv 6997801, Israel.
[3] Applied Physics Division, Soreq NRC, Yavne 81800, Israel
[4] AFEKA - Tel-Aviv Academic College of Engineering, 69107 Tel-Aviv, Israel.

## 1) Comparison between the different plasma models

Thin plasma formations impose a phase on the wavefront of the probe, resulting in the diffraction of the latter. Since the phase is integrated along the transverse direction of the plasma, one must consider the spatial density distribution of the plasma using an appropriate model. In what follows we compare between the proposed Supergaussian density distribution model (SG) with the commonly used pure Gaussian and uniform cylindrical models. We examine two scenarios: A diffraction pattern from low-density plasma, generated by relatively low pulse energy of $200\mu J$ (Fig.S1a), and from higher density plasma generated by higher pulse energy of $960\mu J$ (Fig.S1b).

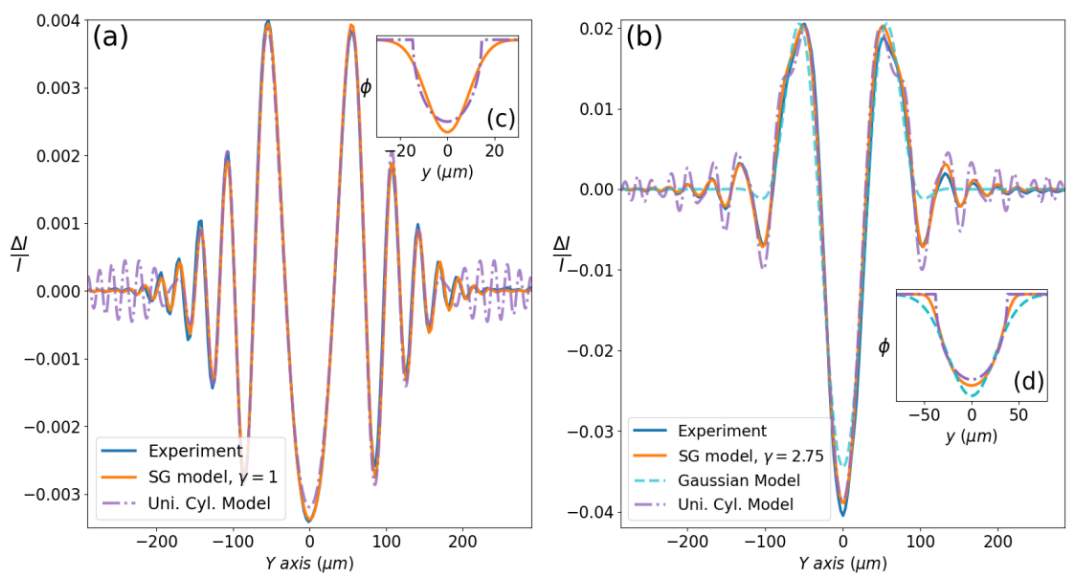

Figure S1: Comparisons between the SG plasma model proposed here and the commonly used Gaussian and uniform cylinder models. (a) Diffraction pattern from low-density plasma formation induced by 200μJ pulse energy. The experimental diffraction pattern is depicted in blue, uniform cylinder model in purple and the SG model with $\gamma = 1$ in orange. (b) Diffraction pattern from plasma formation induced by 960μJ pulse energy. (c,d) Plasma phases retrieved from the different fitted models form Fig.(a) and (b) respectively.

The experimentally measured diffraction patterns are depicted by the blue curves in Fig.S1. The phase derived from the uniform cylinder model and the pure Gaussian model are depicted by the purple and pale blue curves and the phase derived from the SG model is depicted by the orange curve. In Fig.S1a we see that the diffraction from weakly ionized plasma is perfectly captured by the pure Gaussian model (orange curve). In fact, we fitted the extracted phase to a SG plasma distribution with $\gamma$ as a free parameter and obtained $\gamma = 1$ (pure Gaussian distribution). The failure of the uniform cylindrical model is readily observed (dash-dotted purple curve). The phase derived from the latter (shown in Fig.S2c) exhibits sharp discontinuity in its derivative at the edges, that lead to diffraction features resembling those of a wire, where a 'sinc'-like modulations extend at the sides of the diffraction pattern. Thus, for low density plasma formations the SG model converges to a pure Gaussian distribution with $\gamma = 1$ and performs well, while the uniform cylinder model fails. The resulting plasma parameters obtained from a fit to the uniform cylinder model are $n_e = 10^{17} cm^{-3}, r = 14.5 \mu m$ whereas the supergaussian model yields $\rho_0 = 1.53 \cdot 10^{17} cm^{-3}, \sigma = 9 \mu m, \gamma = 1$. This translates to a discrepancy of ~25% for the plasma density $n_e$ and ~36% for the object size (by comparing FWHM calculated from the plasma parameters noted above).

In Fig.S2b we compare the fitting quality of the three models to the experimental diffraction pattern for denser plasma formation. The superiority of the supergaussian model is readily evident by its excellent fit quality compared to the two other models. The experimental diffraction pattern exhibits fast oscillations observe in Fig.S1b starting from $y \pm 130 \mu m \rightarrow \pm 200 \mu m$ respectively. Those are a direct consequence of the abrupt discontinuity in the phase, as shown in Fig.S1d. While the uniform cylindrical model (purple curve) naturally exhibits these fast oscillations, their magnitude and overall correspondence with the experimentally retrieved phase is rather poor. The pure Gaussian model completely lacks the oscillatory diffraction signals rendering its agreement with the experimental results very poor.

The resulting plasma parameters obtained from a fit to the uniform cylinder model are $n_e = 2.6 \cdot 10^{17} cm^{-3}, r = 38 \mu m$, for the supergaussian model $n_e = 2.9 \cdot 10^{17} cm^{-3}, \sigma = 28 \mu m, \gamma = 2.75$ and $n_e = 3.8 \cdot 10^{17} cm^{-3}, \sigma = 25 \mu m$ for the Gaussian model. This translates to significant discrepancies of ~30% in plasma density ($n_e$) and ~13% in plasma width.

Figure S2 depicts a set of supergaussian distributions given by $a \cdot e^{-\left(\frac{r}{\sqrt{2}\sigma}\right)^{2\gamma}}$ where $a = 1$, $\sigma = 20 \mu m$ and varying $\gamma$. For $\gamma = 1$, the distribution is purely Gaussian and as $\gamma$ increases, the distribution gradually coincides with the uniform cylinder model, thus the supergaussian effectively bridges between the two commonly used models in a smooth and continuous way.

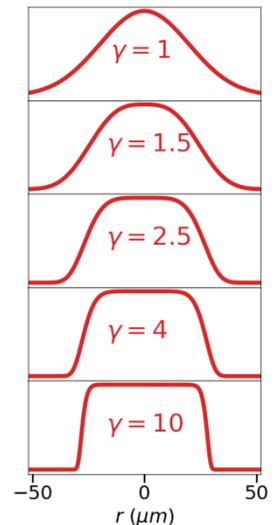

Figure S2: SG distributions with varying SG order parameter $\gamma$.

## 2) Examples of measured diffraction patterns and their fitted SG distributions

Figures S3a-c depict three diffraction patterns out of many measured in our experiment at different positions along the plasma formation (purple data points). Note the variance in modulation depth from $4\%$ in Fig.S3a down to $0.1\%$ in Fig.S3c. Analytically extracted phases from the three diffraction patterns are shown in Figs. S3d-f respectively (blue data points) and are fitted by the SG distribution model (orange curve). The SG plasma parameters $\rho_0, \sigma, \gamma$ are noted in the figures. Next, we simulated the three plasma distributions using the extracted parameters, performed 1D integration and extracted the plasma phase that was imposed on a simulated probe field and propagated using the Fresnel diffraction formalism. The resulted diffraction patterns were overlaid on Figs.S3(a-c) in Cyan to validate our analysis procedure. The pulse was focused by a $f = 500mm$ lens, diffraction was measured $25ps$ past plasma generation and the screen (PD) was situated at a distance of $5.3mm$ from the plasma position.

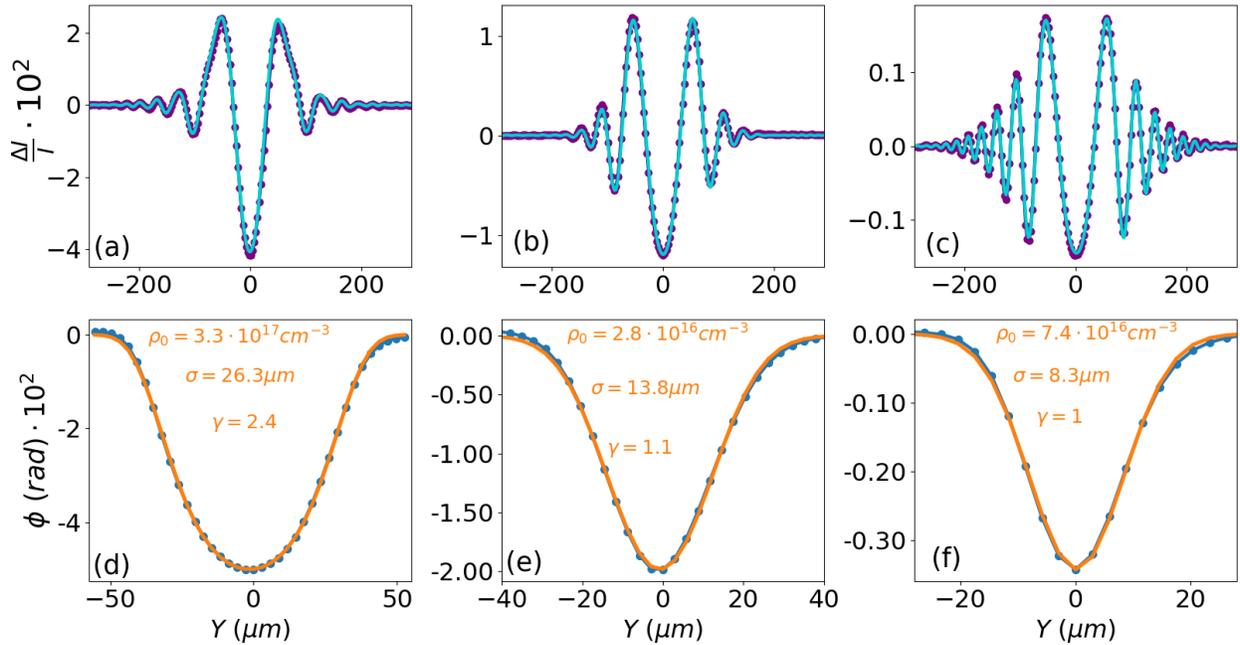

## 3) Identifying the geometric focal point

The experimental results presented in the main article file (Fig.3) show the evolution of the plasma parameters along the z-propagation axis. In order to accurately determine the focal point of the beam (marked $z = 500mm$) we mounted a thin $80\mu m$ glass slide at the point of intersection between the probe pulse and the pump pulse with the probe blocked. The pump beam was attenuated to induce minimal (yet detectable) Self-Phase Modulation upon focusing at the thin glass surface. The SPM was characterized by the spectral modifications detected by a spectrometer placed at the output from the glass. The position of the lens (mounted on translation stage S1) was scanned. With this method we located the geometric focal point (of the lens) with accuracy better than $200\mu m$. To further validate the position of the focal point, we have measured the position of the lowest detectable diffraction pattern signal as a function of the position of the lens (with the glass removed).

## 4) Simulation of the Non Linear Schordinger Equation (NLSE)

In order to simulate the non-linear pulse evolution and plasma formation we the propagate the Non-Linear Schrödinger Equation (NLSE) following the procedure described in ref [1] and [2] :

$$\frac{\partial E}{\partial z} = \frac{i}{2k_0}\nabla_T^2 E - \frac{ik''}{2}\frac{\partial^2 E}{\partial t^2} + ik_0 n_2 |E|^2 E - \frac{\sigma}{2}(1+i\omega_0\tau_c)\rho E - \frac{\beta_k}{2}|E|^{2K-2}\left(1-\frac{\rho}{\rho_{nt}}\right)E$$

The terms on the right hand side include the effects of diffraction, group velocity dispersion (GVD), Kerr effect, plasma absorption and defocusing, and multi photon ionization losses.

$k_0$ is the wavenumber, $\nabla_T^2 = \frac{\partial^2}{\partial x^2} + \frac{\partial^2}{\partial y^2}$, $k'' = 0.2 fs^2 cm^{-1}$ is the GVD, Kerr coefficient $n_2 = 3 \cdot 10^{-19} W^{-1} cm^2$, cross section for inverse Bremsstrahlung $\sigma = 5.6 \cdot 10^{-20} cm^2$, $\tau_c = 350 fs$ is the electron collision time, $\rho$ is the plasma density, density of neutrals ($O_2$ molecules only), $\rho_{nt} = 5.4 \cdot 10^{18} cm^{-3}$, $\beta_k = K\hbar\omega_0\rho_{nt}\sigma_k = 3 \cdot 10^{-95} W^{-7} cm^{13}$, ionization potential of $O_2$, $U_i = 12.063 eV$. 8 photon ionization with $K = 8$ was assumed, ionization cross section $\sigma_k = 2.81 \cdot 10^{-96} W^{-8} cm^{16} s^{-1}$.

For propagating the equation numerically we used the split-step method [2]. To reduce computation time and resources, the beam is assumed radially symmetric and the problem was solved in (2D+1) dimensions. The evolution of the pulse is tracked along the propagation axis Z in both *t* and *r*. Convergence in each parameter was tested and verified.

Each propagation step is split into three sub-steps:

1) Propagation in spatial domain by solving $\frac{\partial E}{\partial z} = \frac{i}{2k_0}\nabla_T^2 E$. This is done by transposing the problem to the spatial frequency domain using Hankel Transform (see https://github.com/etfrogers/pyhank). This translates to propagating as $\tilde{E}_{r,n+\frac{1}{3}} = \tilde{E}_{r,n} e^{-i\frac{k_T^2}{2k}\Delta z}$.

2) Propagation in the time domain using FFT algorithm, similarly to (1): $\tilde{E}_{t,n+\frac{2}{3}} = \tilde{E}_{t,n} e^{i\frac{k''\omega}{2}\Delta z}$.

3) Multiplying by the local/temporal non-linear terms: $E_{n+1} = E_{n+\frac{2}{3}} e^{N\Delta z}$
   Where $N = ik_0 n_2 I^2 + \frac{\sigma}{2}(1+i\omega_0\tau_c)\rho - \frac{\beta_k}{2}|E|^{2K-2}\left(1-\frac{\rho}{\rho_{nt}}\right)$

The step size $\Delta z$ was chosen dynamically, i.e. $\Delta z$ was allowed to vary over the course of propagation. The step size in z,t,r were restricted to: $\Delta z_{min} = 20\mu m$, $\min \Delta t_{min} = 3 fs$, $\Delta r_{min} = 1.7 \mu m$. The simulation was checked for convergence in all (2D+1) dimensions.

The plasma density was calculated in each step according to:

$$\rho(t_n) = \rho(t_{n-1}) + \Delta\rho(t_n)$$

$$\Delta\rho(t_n) = \sigma_k \frac{I^K(t_n)+I^K(t_{n-1})}{2}(\rho_{nt}-\rho(t_{n-1})) + \frac{\sigma}{U_i}\rho(t_{n-1})\frac{I(t_n)+I(t_{n-1})}{2}$$

In order to create figure 3 of the main text file, we fitted each slice of the plasma profile to a SG function and extracted the plasma parameters $\rho_0, \sigma, \gamma$ at time $300 fs$ (after passage of the pulse). The SG plasma distribution was found in excellent agreement with the simulated results, and as exemplified in Fig.S2 for three representative diffraction patterns.

A simplified 'Frozen time' [3] [4] version of the simulation where the temporal domain is discarded from the dynamics was utilized to extract the intensity and phase profiles of Fig.4 in the main text file. This was done by replacing $\rho(r,t)$ with $\rho \approx 0.376 \cdot \sigma_k \tau \rho_{nt} I^k(r)$. The 0.376 factor stems from integration of the pulse intensity over the pulse duration. This simplified treatment discards any modifications imposed on the pulse along its propagation (such as pulse splitting, broadening or shortening) and discards the effects of temporal dispersion. We note that for low pulse energies, both simulations predict similar results with no significant effects of associated with temporal modifications of the pulse.

## 5) The effect of finite slit width on the extracted $\gamma$ parameter

What may seem as a discrepancy between the experimental and theoretical SG order parameter $\gamma$ is found to be nothing else but the convolution of the diffraction signal with the window function of the $20\mu m$ slit that is mounted on the detection PD. In Fig.2 of the main text file we showed the plasma parameters that were extracted from the raw diffraction pattern (as measured). The theoretical plasma parameter of Fig.3 were extracted by direct fitting of the simulation results to SG distribution. In Figure S4 we simulate two diffraction patterns expected from plasma parameters that were obtained from the NLSE calculation (blue curves in Fig.a,c). To simulate the experimental diffraction patterns (Orange curves in Figs.a,c) we convolve the blue signals with the window function of the slit. We then extract the SG phase and plasma parameters using the analytical extraction scheme discussed in the main text (Figs.b,d with same color coding). Evidently the convolution does not affects the density $\rho_0$ and width σ plasma parameters while the γ parameter is highly sensitive with $\gamma$ varying from $6.03 \rightarrow 2.62$ and $2.12 \rightarrow 1.34$ due to the convolution with the window function of the slit. Note that this relieves the discrepancy and validates the excellent agreement between the experimental and theoretical plasma parameters as reported in the text.

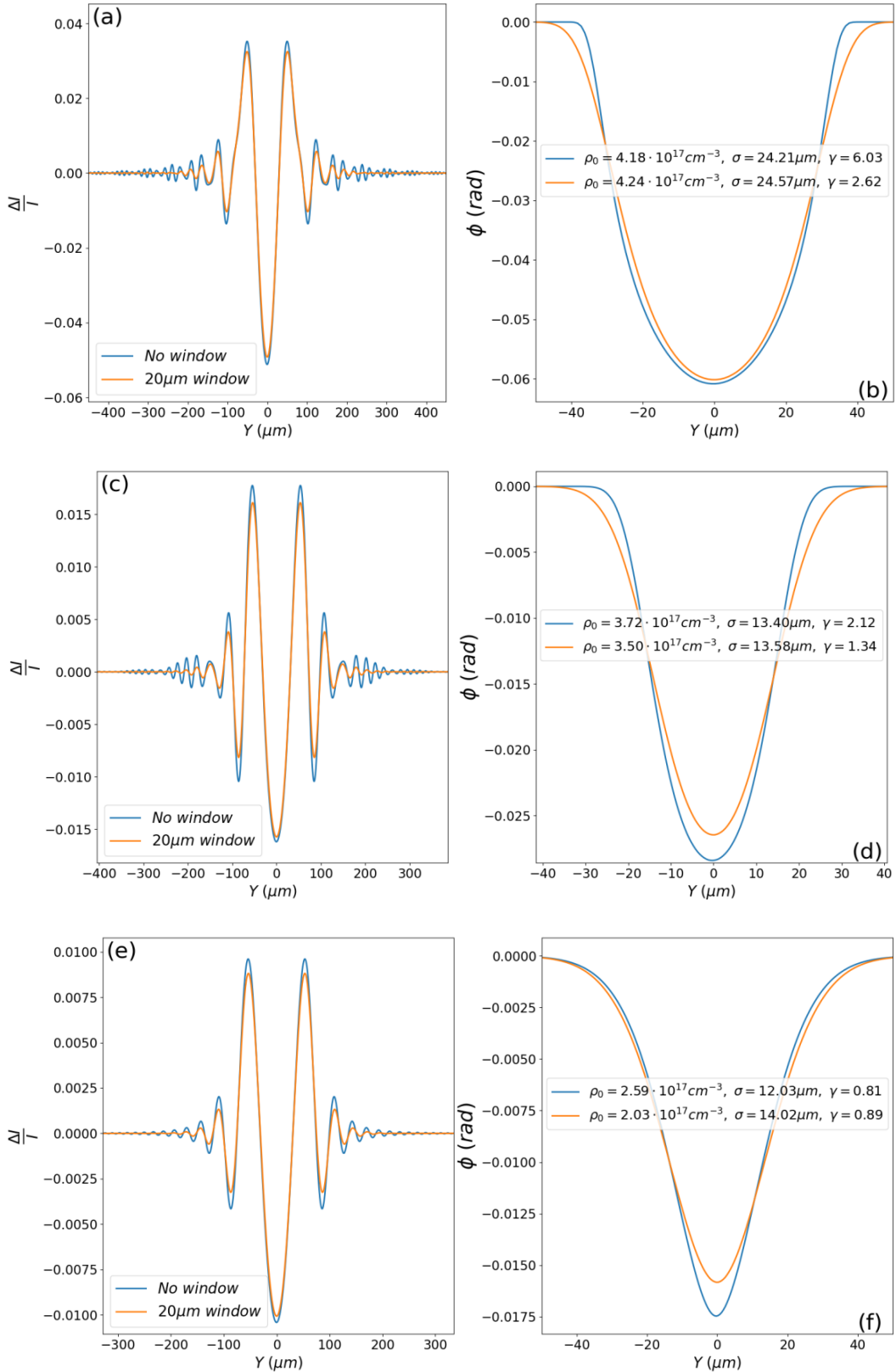

Figure S4: (a,c,e): Simulated diffraction patterns and (b,d,f) their corresponding SG fitting parameters. The 'as simulated' diffraction patterns are depicted in blue while the diffraction patterns following convolution with the $20\mu m$ window function of the slit are depicted in orange.